# Gate tuning of electronic phase transitions in two-dimensional NbSe$_2$

Xiaoxiang Xi[1], Helmuth Berger[2], László Forró[2], Jie Shan[1]*, and Kin Fai Mak[1]*

[1]Department of Physics and Center for 2-Dimensional and Layered Materials, The Pennsylvania State University, University Park, Pennsylvania 16802-6300, USA

[2]Institute of Condensed Matter Physics, Ecole Polytechnique Fédérale de Lausanne, 1015 Lausanne, Switzerland

*Correspondence to: jus59@psu.edu, kzm11@psu.edu

## Abstract

Recent experimental advances in atomically thin transition metal dichalcogenide (TMD) metals have unveiled a range of interesting phenomena including the coexistence of charge-density-wave (CDW) order and superconductivity down to the monolayer limit. The atomic thickness of two-dimensional (2D) TMD metals also opens up the possibility for control of these electronic phase transitions by electrostatic gating. Here we demonstrate reversible tuning of superconductivity and CDW order in model 2D TMD metal NbSe$_2$ by an ionic liquid gate. A variation up to ~ 50% in the superconducting transition temperature has been observed, accompanied by a correlated evolution of the CDW order. We find that the doping dependence of the superconducting and CDW phase transition in 2D NbSe$_2$ can be understood by a varying electron-phonon coupling strength induced by the gate-modulated carrier density and the electronic density of states near the Fermi surface.

In metals, the interactions between free carriers and ions that form the crystal structure lead to a multitude of many-body electronic phases [1]. Modifying the free carrier density, and thus the density of states (DOS) near the Fermi surface and screening of the interaction effects, has long been sought to control the electronic phases [2]. Electrostatic gating is a clean and reversible method to introduce doping near the surface of a material [3]. In particular, electric-double-layer based on an ionic liquid has had a tremendous success in doping various insulator surfaces into metallic phases [4-15]. However, little success has been made on significantly changing the carrier density in metals by electrostatic gating due to screening arisen from the extremely high carrier densities [16-18]. Recent experimental advances in atomically thin transition metal dichalcogenide (TMD) metals have unveiled a range of interesting collective electronic phenomena, including the coexistence of charge-density-wave (CDW) order and superconductivity down to the monolayer limit [19-21], a Bose metal phase [22], and Ising pairing in superconductivity [23]. The atomic thickness of two-dimensional (2D) TMD metals has also provided an ideal system to explore electrostatic doping and control of the collective electronic phases in metals.

2H-NbSe$_2$ is a representative TMD metal made of layers bonded by van der Waals' interactions [24]. Each single NbSe$_2$ layer (half of a unit cell thickness) consists of a layer of transition metal Nb atoms sandwiched between two layers of chalcogen Se atoms, forming a trigonal prismatic structure. The electronic band structure near the Fermi energy shows multiple Fermi pockets formed by the valence Bloch states (figure 1a for bilayer NbSe$_2$) [25]. 2H-NbSe$_2$ is thus a hole metal at high temperature. It undergoes a CDW and superconducting transition, respectively, at ~ 33 and 7 K in the bulk, with the two collective phases coexisting below 7 K [24]. Because of the weak interlayer van der Waals' bonding, NbSe$_2$ has been successfully exfoliated into atomically thin layers [19,20,22,23,26,27]. Monolayer NbSe$_2$ has also been grown on graphene by molecular beam epitaxy (MBE) [21]. With decreasing layer thickness, whereas all studies consistently show a decreasing superconducting transition temperature $T_C$ (detailed values vary among samples of different origin) [19-23], a large discrepancy exists in the reported CDW transition temperature $T_{CDW}$. For instance, Xi et al. observed strongly enhanced CDW order in atomically thin NbSe$_2$ with $T_{CDW} > 100$ K for exfoliated monolayer samples [19]. Ugeda et al. reported slightly weakened CDW order ($T_{CDW}$ ~ 25 K) in monolayer NbSe$_2$ grown on graphene [21]. These discrepancies are not well understood although sample quality and substrates are believed to play a role.

In this Letter, we employ transport and magnetotransport measurements to investigate the superconducting and CDW phase transition in 2D NbSe$_2$ as a function of carrier density modulated by ionic liquid gating. We have been able to *reversibly* tune the carrier density in bilayer NbSe$_2$ up to $6 \times 10^{14}$ cm$^{-2}$ (~ 30% of the intrinsic density), and the superconducting transition temperature by ~ 50%, with $T_C$ enhanced with hole doping. Although $T_{CDW}$ cannot be accurately determined, similar trend of the CDW transition with doping has been observed. The observed doping dependence of the superconducting and CDW transition can be understood as a result of electron-phonon (e-ph) coupling modulated by the carrier density and the corresponding electronic DOS near the Fermi surface. Our results thus open a new possibility of continuously tuning and studying the electronic phase diagrams of 2D metals over a large window of doping density by ionic liquid gating.

Atomically thin $NbSe_2$ is known to be unstable under ambient conditions [19,20,22,23,27]. In fact, recent advances on the study of its intrinsic properties have been made possible only by capping $NbSe_2$ [19,20,22,23] or performing in-situ measurements [21]. In a recent gating study of thick $NbSe_2$ flakes using an ionic liquid [28], $NbSe_2$ flakes were brought into direct contact with the ionic liquid and an irreversible behavior was observed. This is presumably caused by electrochemistry on the surface of $NbSe_2$. To protect the 2D $NbSe_2$ samples from oxidation and undesired electrochemistry in ionic liquid gating while maintaining the high gate capacitance of the device, we have introduced an ultrathin capping layer of chemically stable large-gap semiconductors [29,30] or insulators [13,14]. Both $MoS_2$ monolayers and ultrathin (~ 1 nm) hexagonal boron nitride (hBN) layers have been tested. Within the gate voltage range of interest, they have produced similar results (Supplementary Materials Section 1). In this work, we have mainly used monolayer $MoS_2$ to cap $NbSe_2$ because it is much easier to identify than ultrathin hBN due to its larger optical contrast in the visible.

The fabrication method of 2D $NbSe_2$ devices has been described elsewhere [23]. In brief, atomically thin $NbSe_2$ samples were mechanically exfoliated from bulk single crystals on silicone elastomer polydimethylsiloxane (PDMS). Their thickness was first identified by the optical reflection contrast and later confirmed by Raman spectroscopy [19]. The samples were then transferred onto silicon substrates with pre-patterned Au electrodes and shaped into a Hall bar geometry by removing unwanted areas using a polypropylene carbonate (PPC) layer on PDMS stamps. Capping layers (monolayer $MoS_2$ or ~ 1 nm thick hBN), which were prepared on separate substrates, were then transferred onto $NbSe_2$. Ionic liquid *N*,*N*-diethyl-*N*-(2-methoxyethyl)-*N*-methylammonium bis(trifluoromethylsulphonyl-imide) (DEME-TFSI) was finally drop casted, covering both the sample and the gate electrode. The finished devices were annealed at 350 K in high vacuum for 3 – 5 hours to dehydrate the ionic liquid and to ensure good interfacial contacts between the different layers. The schematic and optical microscope image of a typical device are shown in figure 1b & 1c, respectively. Compared to $NbSe_2$ monolayers, it is much easier to fabricate high quality bilayer samples and devices. Below we focus our study on $NbSe_2$ bilayers. Similar results but stronger effects are expected for monolayers.

Transport measurements were carried out in a Physical Property Measurement System down to 2.1 K. Both longitudinal sheet resistance ($R_s$) and transverse sheet resistance ($R_t$) were acquired with excitation currents limited to 1 µA to avoid heating and high-bias effects. Because of the finite longitudinal-transverse coupling in our devices and the presence of magnetoresistance at low temperature, we antisymmetrized the transverse sheet resistance under magnetic field $H$ of opposite directions to obtain the Hall resistance, $R_{xy}(H) = \frac{R_t(H)-R_t(-H)}{2}$, and the sheet Hall coefficient, $R_H = \frac{R_{xy}}{H}$. (See Supplementary Section 2 for more details.) To vary the doping density, gate voltage was adjusted at 220 K (which is near the freezing point of the ionic liquid [8]), followed by rapid cooling to minimize any electrochemistry effects.

Figure 2 shows the temperature dependence of the sheet resistance $R_s$ and the sheet Hall coefficient $R_H$ of a typical device (#120) under gate voltage $V_G$ varying from -2 V to 3 V. Metallic behavior is seen in the temperature dependence of $R_s$ (figure 2a): $R_s$ scales linearly with $T$ due to e-ph scattering at high temperature, and saturates to a residual value of about 250 Ω due to impurity/defect scattering at low temperature. The residual-resistance ratio (RRR) (estimated as $\frac{R_s(300K)}{R_s(8K)}$) is about 6. Further cooling causes a rapid drop of $R_s$ to zero below 5 K, indicating the superconducting transition. Fig. 2b clearly shows that $T_C$ shifts by ~ 30% under gate voltage varying from -2 V to 3 V. This modification is in contrast to previous experiments based on silicon or oxide back gates, where very small modulations in $T_C$ (< 0.2 K) have been observed [20,27,31]. Similarly, $R_H$ also depends strongly on gate voltage (figure 2c): for $T > 100$ K, $R_H$ is largely temperature independent; its value increases with increasing $V_G$. For $T < 100$ K, a drop in $R_H$ upon cooling is observed under $V_G$ up to ~ 2 V; the drop in $R_H$ disappears under large positive $V_G$'s. As we discuss below, the high-temperature behavior of $R_H$ is dominated by the carrier density and will be used to evaluate its value; and the low-temperature behavior is influenced by the CDW transition.

Before we interpret the above gating effects on the electronic phase transitions, multiple control experiments have been performed. (See Supplementary Section 1 for details.) We first test the reversibility of the effect. Six bilayer devices have been tested under different gating sequences. All of them were highly reversible and repeatable under the gate voltage range of -2 V to 3 V. (Some devices withstood higher gate voltages, e.g. #150.) Thus extrinsic effects such as gate-induced electrochemistry are unlikely the origin of the observed effects. Second, we compare ionic liquid gating with conventional solid-state dielectric gating by fabricating a bilayer NbSe$_2$ device with a combination of a Si/SiO$_2$ back gate and an hBN/graphene top gate. Similar effects on $T_C$, but with much smaller modulations, have been observed in the latter device. Third, we verify the role of the MoS$_2$ capping layer. To exclude the possibility of gate-induced superconductivity in MoS$_2$, as has been recently demonstrated [8,10,11,15], we have performed an identical experiment on monolayer MoS$_2$ alone. No gate-induced superconductivity was observed within the gate voltage range employed in this work. Moreover, electrical contacts to monolayer MoS$_2$, prepared by our fabrication method, are highly unstable and often exhibit zero conductance (i.e. doping into MoS$_2$ is unlikely). The MoS$_2$ film thus only serves as a protection dielectric layer similar to ultrathin hBN. We also note that in addition to doping, electrostatic gating with a single gate introduces a vertical electric field on the samples. The device with dual solid-state gates showed that the electric-field effect on the electronic phase transitions is negligible compared to the doping effect. We therefore conclude that the observed gating effects on the transport measurements arise primarily from modulations in collective electronic phases induced by doping in NbSe₂.

We now extract the doping dependence of the superconducting transition temperature of bilayer NbSe₂. For simplicity, we have taken $T_C$ to be the temperature corresponding to half of the normal state resistance (see Supplementary Section 3 for more accurate determinations of $T_C$ and for current excitation measurements at varying temperatures). To extract the total sheet density $n_{2D}$, we have used the high-temperature (> 100 K) value of the Hall coefficient $R_H$. Although NbSe₂ is a multiband metal, it has been shown that at high temperature the carrier scattering rate becomes isotropic in the Fermi surface and $R_H$ can provide a good estimate of the carrier density by using $n_{2D} = \frac{f}{eR_H}$ [32]. Here $e$ is the elementary charge, and $f$ is dimensionless and is close to unity. A value of $f \approx 0.8$ has been obtained by calibrating $n_{2D}$ at $V_G = 0$ V to the known carrier density $n_0$ in bulk NbSe₂: $n_{2D} = n_0 t \approx 1.9 \times 10^{15} cm^{-2}$ ( $n_0 \approx 1.5 \times 10^{22} cm^{-3}$ corresponding to 1 hole per Nb atom and $t \approx 1.25$ nm for bilayer NbSe₂ [24]). We note that in the above calibration we have ignored charge transfer between MoS₂ and NbSe₂ due to their work function mismatch, which has been estimated not to exceed $10^{13}$ cm⁻². The $V_G$-dependence of $n_{2D}$ is shown in figure 2d (symbols), which follows a linear dependence (dashed line). The negative slope is consistent with the fact that NbSe₂ is a hole metal. We calculate the gate capacitance from the slope to be 7 $\mu F/cm^2$. The value agrees very well with the reported ones for the same ionic liquid [8]. The corresponding tuning range of the Fermi energy, calibrated from the DOS at the Fermi energy from *ab initio* calculations [25], is ~ 130 meV (shaded region in the electronic band structure of figure 1a). Figure 3 summarizes the doping dependence of $T_C$ for three devices. It clearly shows a monotonic dependence of $T_C$ on the (hole) density in 2D NbSe₂ and the ability of gate tuning of $T_C$ from 2.8 K to 5.2 K (4.5 K for undoped case) corresponding to a modulation of ~ 50%.

What is the microscopic origin of the doping dependence of $T_C$ ? Superconductivity in bulk NbSe₂ is known to be the BCS type driven by e-ph interactions. Indeed, the doping-induced change in $T_C$ is correlated with the change in the e-ph interactions in our bilayer devices. The slope of the temperature dependence of the resistance ($\frac{dR_S}{dT}$) at high temperature, indicating the e-ph scattering strength, can be seen to increase with hole doping (Fig. 2a). Below we employ a simplified model to relate $T_C$ to doping. The model, which has been applied successfully in previous studies on bulk NbSe₂ [33], ignores the multiband nature of NbSe₂. In our model, $T_C$ is predicted by the strong-coupling formula as [34]

$$T_C = \frac{\omega_{log}}{1.2}\exp[-\frac{1.04(1+\lambda)}{\lambda-\mu^*(1+0.62\lambda)}]. \tag{1}$$

Here $\omega_{log}$ is the weighted average of the phonon energies in Kelvin introduced by Allen and Dynes [34], $\mu^*$ is the Coulomb pseudopotential, and $\lambda = N(\epsilon_F)V_0$ is the dimensionless e-ph coupling constant with $N(\epsilon_F)$ and $V_0$ denoting, respectively, the electronic DOS at the Fermi energy $\epsilon_F$ and the effective e-ph coupling energy $V_0$. We evaluate the e-ph coupling constant for each carrier density $n_{2D}$ (figure 4a) from $\frac{dR_s}{dT}$ at high temperature by considering electron-phonon scattering [1] for a 2D metal: $\frac{dR_s}{dT} = \frac{2\pi\hbar n_0 k_B}{\varepsilon_0 n_{2D}(\hbar\omega_{p0})^2}\lambda$ (See Supplementary Section 4 for details). Here $\hbar$, $k_B$, $\varepsilon_0$ and $\hbar\omega_{p0}(\approx 2.74$ eV [35]) are the Planck's constant, Boltzmann constant, vacuum permittivity, and the in-plane plasma energy of bulk NbSe$_2$, respectively. The dependence of $T_C$ on $\lambda$ is shown in Fig. 4b (symbols), which can be reasonably well described by Eq. (1) with $\omega_{log} \approx 50$ K and $\mu^* \approx 0.10$ (solid line). The value of $\omega_{log}$ agrees well with the reported one from the layer thickness dependence of $T_C$ for 2D NbSe$_2$ [23]. The value of $\mu^*$ is also consistent with the estimated value of $\mu^* \approx \frac{0.26 N(\epsilon_F)}{1+N(\epsilon_F)} \approx 0.15$ [36] using the DOS at the Fermi energy from *ab initio* calculations for undoped NbSe$_2$ [25]. Future studies on 2D NbSe$_2$ or similar systems by introducing higher doping densities so that the Fermi level approaches the saddle point singularity at the M-point of the Brillouin zone [25] would be very interesting. It may lead to drastic changes and new phenomena in the collective electronic phases such as a superconducting dome which has been observed in heavily doped TMD MoS$_2$ [8].

Finally, we comment on the CDW order in 2D NbSe$_2$. Signature of CDW in the temperature dependence of the resistance as a kink around $T_{CDW}$ has been observed in bulk NbSe$_2$ samples with large RRRs [37]. No kink can be identified here for atomically thin NbSe$_2$ samples (Fig. 2a) presumably due to their low RRRs. Furthermore, the strong inelastic light scattering from the ionic liquid prevents us from determining $T_{CDW}$ in atomically thin NbSe$_2$ by Raman spectroscopy [19]. Here we use the temperature dependence of the Hall coefficient to evaluate the CDW order. It has been shown in bulk NbSe$_2$ that a significant drop in the Hall resistance upon cooling occurs near $T_{CDW}$ due to Fermi surface instability [37,38]. (The Hall resistance changes sign near $T_{CDW}$ if the sample's RRR > 25 [37].) We observed a similar behavior in bilayer NbSe$_2$ for $V_G$ up to ~ 2 V (figure 2c). At $V_G = 3$ V, the drop in $R_H$ disappears. Instead it increases monotonically with decreasing *T*, which resembles the behavior of bulk NbS$_2$ [33] (empty triangles, figure 2c). NbS$_2$ has similar electronic properties as NbSe$_2$, but does not possess a CDW phase [24,33]. These observations suggest that the CDW order in bilayer NbSe$_2$ is weakened with increasing $V_G$ (decreasing hole density) and can be destroyed at large gate voltages. The positive correlation between superconductivity and the CDW order here contrasts from the weak anti-correlation observed earlier in bulk NbSe$_2$ under high pressure [39], in which a slight increase in $T_C$ has been observed under complete destruction of the CDW order. Recent angle-resolved photoemission spectroscopy (ARPES) [40] shows that while anisotropic s-wave superconducting gaps are opened at the NbSe$_2$ Fermi surface, CDW gaps are opened only near the CDW wavevectors, where the superconducting gap is minimum; complete destruction of the CDW gaps only slightly enlarges the Fermi surface for superconductivity. Our results are consistent with the fact that both superconductivity and the CDW order are driven by e-ph interactions. Hole doping increases the electronic DOS near the Fermi surface, which leads to stronger e-ph interactions and stronger superconductivity and CDW order.

**References**

[1] G. Grimvall, *The Electron-Phonon Interaction in Metals* (North-Holland Publishing Co., 1981).
[2] R. E. Glover and M. D. Sherrill, Physical Review Letters **5**, 248 (1960).
[3] C. H. Ahn *et al.*, Reviews of Modern Physics **78**, 1185 (2006).
[4] K. Ueno, S. Nakamura, H. Shimotani, A. Ohtomo, N. Kimura, T. Nojima, H. Aoki, Y. Iwasa, and M. Kawasaki, Nat Mater **7**, 855 (2008).
[5] A. T. Bollinger, G. Dubuis, J. Yoon, D. Pavuna, J. Misewich, and I. Bozovic, Nature **472**, 458 (2011).
[6] Y. Lee, C. Clement, J. Hellerstedt, J. Kinney, L. Kinnischtzke, X. Leng, S. D. Snyder, and A. M. Goldman, Physical Review Letters **106**, 136809 (2011).
[7] X. Leng, J. Garcia-Barriocanal, S. Bose, Y. Lee, and A. M. Goldman, Physical Review Letters **107**, 027001 (2011).
[8] J. T. Ye, Y. J. Zhang, R. Akashi, M. S. Bahramy, R. Arita, and Y. Iwasa, Science **338**, 1193 (2012).
[9] M. Yoshida, Y. Zhang, J. Ye, R. Suzuki, Y. Imai, S. Kimura, A. Fujiwara, and Y. Iwasa, Scientific Reports **4**, 7302 (2014).
[10] Y. Saito *et al.*, Nat Phys **12**, 144 (2016).
[11] J. M. Lu, O. Zheliuk, I. Leermakers, N. F. Q. Yuan, U. Zeitler, K. T. Law, and J. T. Ye, Science **350**, 1353 (2015).
[12] Y. J. Yu *et al.*, Nature Nanotechnology **10**, 270 (2015).
[13] L. J. Li, E. C. T. O'Farrell, K. P. Loh, G. Eda, B. Özyilmaz, and A. H. Castro Neto, Nature **529**, 185 (2016).
[14] P. Gallagher, M. Lee, T. A. Petach, S. W. Stanwyck, J. R. Williams, K. Watanabe, T. Taniguchi, and D. Goldhaber-Gordon, Nat Commun **6**, 6437 (2015).
[15] D. Costanzo, S. Jo, H. Berger, and A. F. Morpurgo, Nat Nano **11**, 339 (2016).
[16] J. Choi, R. Pradheesh, H. Kim, H. Im, Y. Chong, and D.-H. Chae, Applied Physics Letters **105**, 012601 (2014).
[17] M. Sagmeister, U. Brossmann, S. Landgraf, and R. Würschum, Physical Review Letters **96**, 156601 (2006).
[18] T. A. Petach, M. Lee, R. C. Davis, A. Mehta, and D. Goldhaber-Gordon, Physical Review B **90**, 081108 (2014).
[19] X. Xi, L. Zhao, Z. Wang, H. Berger, L. Forró, J. Shan, and K. F. Mak, Nature Nanotechnology **10**, 765 (2015).
[20] Y. Cao *et al.*, Nano Letters **15**, 4914 (2015).
[21] M. M. Ugeda *et al.*, Nat Phys **12**, 92 (2015).
[22] A. W. Tsen *et al.*, Nat Phys **12**, 208 (2016).

[23] X. Xi, Z. Wang, W. Zhao, J.-H. Park, K. T. Law, H. Berger, L. Forro, J. Shan, and K. F. Mak, Nat Phys **12**, 139 (2016).
[24] J. A. Wilson, F. J. Di Salvo, and S. Mahajan, Adv Phys **50**, 1171 (2001).
[25] M. Calandra, I. I. Mazin, and F. Mauri, Physical Review B **80**, 241108 (2009).
[26] R. F. Frindt, Physical Review Letters **28**, 299 (1972).
[27] N. E. Staley, J. Wu, P. Eklund, Y. Liu, L. J. Li, and Z. Xu, Physical Review B **80**, 184505 (2009).
[28] Z. J. Li, B. F. Gao, J. L. Zhao, X. M. Xie, and M. H. Jiang, Superconductor Science and Technology **27**, 015004 (2014).
[29] K. F. Mak, C. Lee, J. Hone, J. Shan, and T. F. Heinz, Physical Review Letters **105** (2010).
[30] A. Splendiani, L. Sun, Y. Zhang, T. Li, J. Kim, C.-Y. Chim, G. Galli, and F. Wang, Nano Letters **10**, 1271 (2010).
[31] S. E.-B. Mohammed, W. Daniel, R. Saverio, B. Geetha, P. Don Mck, and J. B. Simon, Superconductor Science and Technology **26**, 125020 (2013).
[32] N. P. Ong, Physical Review B **43**, 193 (1991).
[33] M. Naito and S. Tanaka, Journal of the Physical Society of Japan **51**, 219 (1982).
[34] P. B. Allen and R. C. Dynes, Physical Review B **12**, 905 (1975).
[35] S. V. Dordevic, D. N. Basov, R. C. Dynes, and E. Bucher, Physical Review B **64**, 161103 (2001).
[36] K. H. Benneman and J. W. Garland, *Superconductivity in d- and f-band metals* (American Institute of Physics, 1972), p.^pp. 103.
[37] L. Li, J. Shen, Z. Xu, and H. Wang, International Journal of Modern Physics B **19**, 275 (2005).
[38] R. Bel, K. Behnia, and H. Berger, Physical Review Letters **91**, 066602 (2003).
[39] C. W. Chu, V. Diatschenko, C. Y. Huang, and F. J. DiSalvo, Physical Review B **15**, 1340 (1977).
[40] D. J. Rahn, S. Hellmann, M. Kalläne, C. Sohrt, T. K. Kim, L. Kipp, and K. Rossnagel, Physical Review B **85**, 224532 (2012).

**Figures and figure captions**

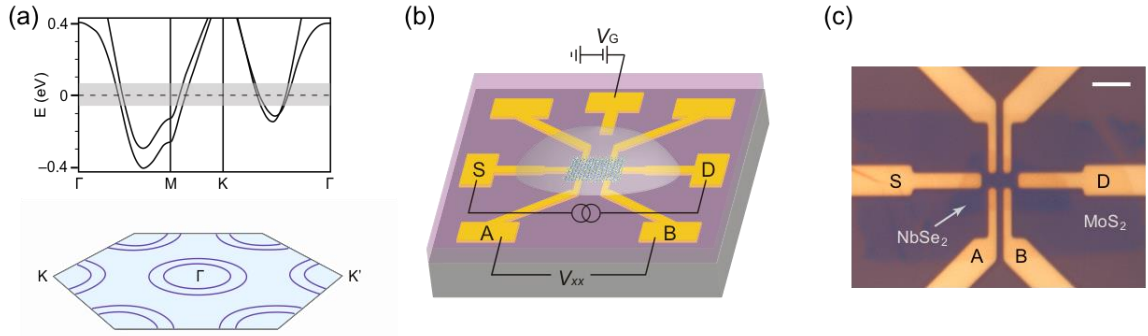

FIG. 1. (a) Top: Electronic band structure of undoped bilayer NbSe$_2$ reproduced from *ab initio* calculations of Ref [25]. The dashed line indicates the Fermi level at zero gate voltage and the shaded region represents the range of Fermi levels accessible by gating in the experiment. Bottom: Schematic of the first Brillouin zone and the Fermi surface around the Γ, K, and K' point. (b) Device schematic: Current was excited through electrode S and D; Longitudinal and transverse voltage drops were measured; Gate voltage $V_G$ was supplied through an isolated electrode from the sample. (c) Optical image of a bilayer NbSe$_2$/monolayer MoS$_2$ stack on Si substrate with gold electrodes before drop casting ionic liquid. Scale bar is 5 μm.

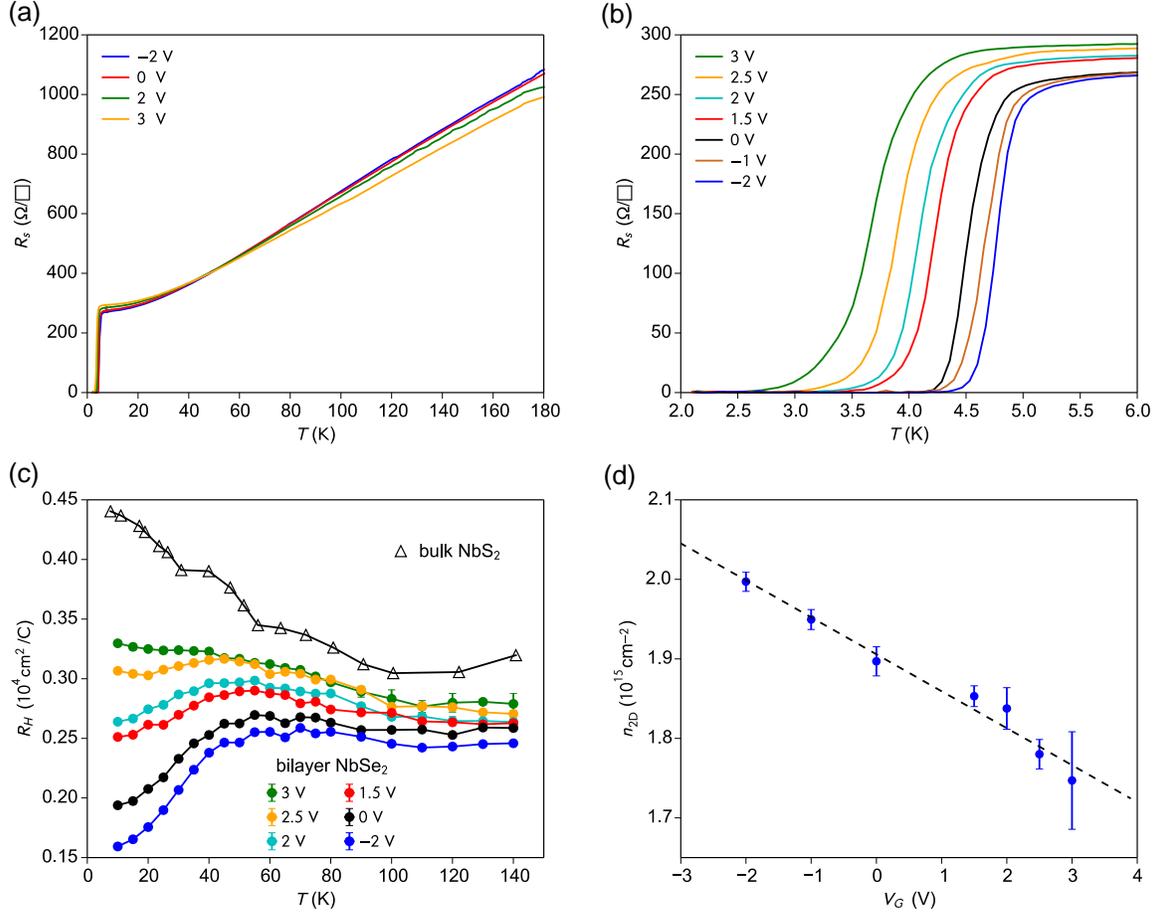

FIG. 2. (a-c) Temperature dependence of the longitudinal sheet resistance (a, b) and the sheet Hall coefficient (c) at selected gate voltages. Longitudinal sheet resistance across the superconducting transition is shown in (b). Data from bulk $NbS_2$ [33] are also shown as empty triangles in (c). (d) Gate voltage dependence of the sheet carrier density. The error bars are estimated from the uncertainties in the sheet Hall coefficient. The dashed line is a linear fit. All data are from device #120.

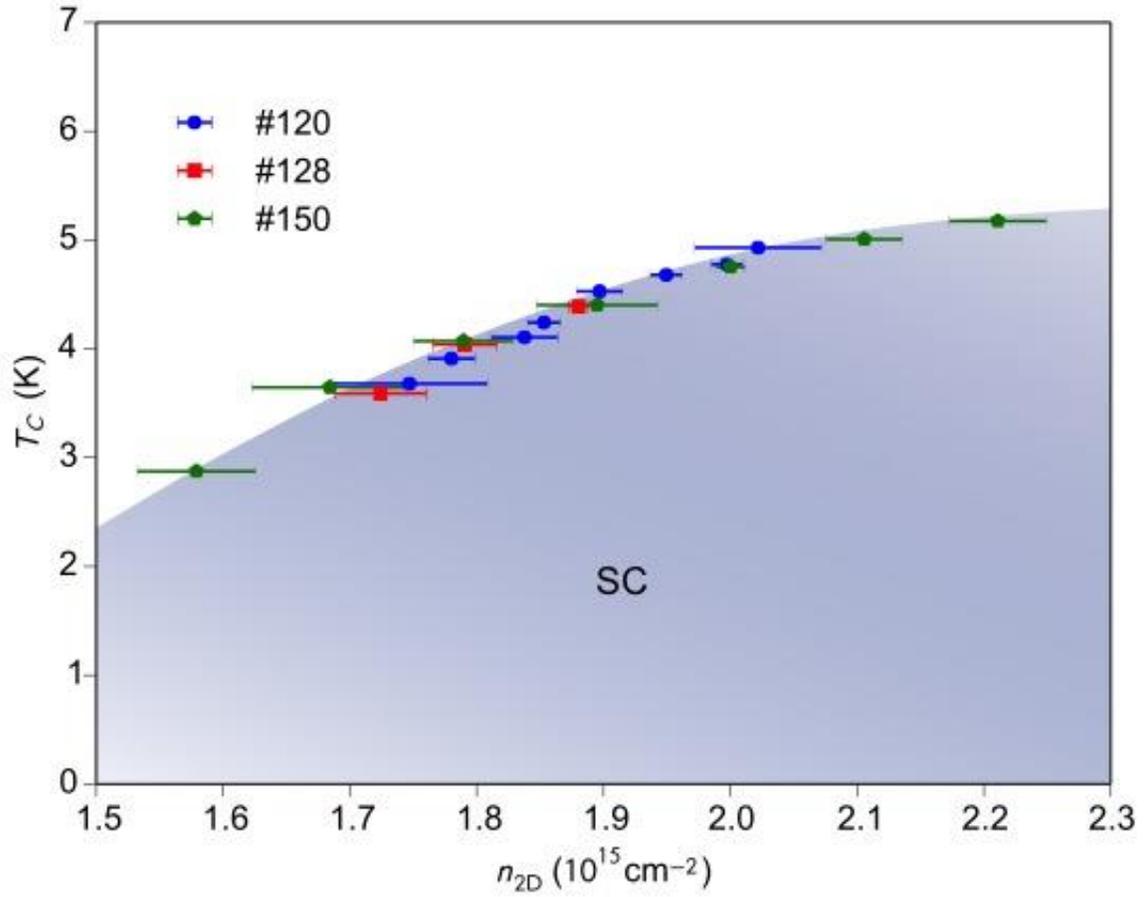

FIG. 3. Superconducting phase diagram of bilayer NbSe$_2$. Different symbols correspond to results from different devices. The horizontal error bars originate from the measurement uncertainty of the Hall coefficient. The filled area corresponding to the superconducting (SC) phase is a guide to the eye.

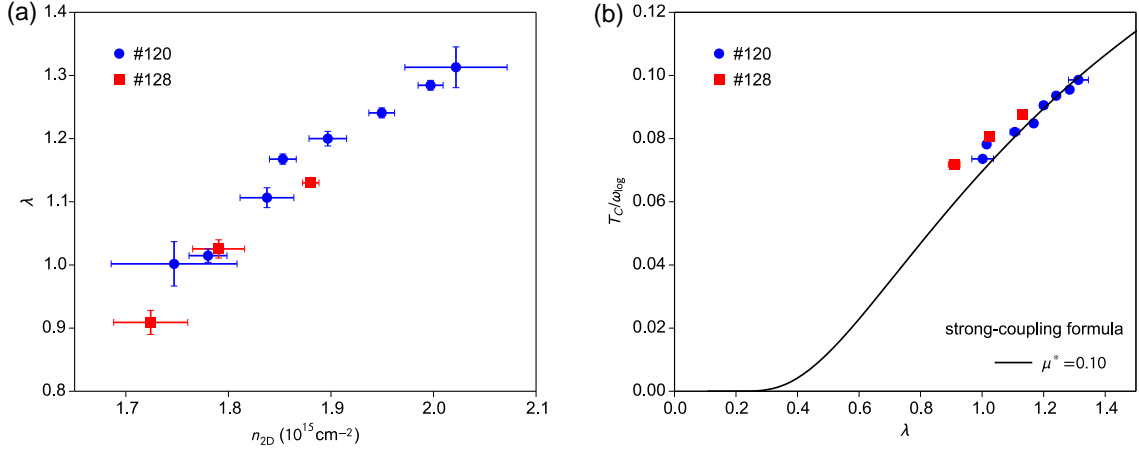

FIG. 4. (a) Dimensionless electron-phonon coupling constant as a function of sheet carrier density. The vertical and horizontal error bars are from fitting of the normal-state resistance to the electron-phonon scattering model and the measurement uncertainty of the Hall coefficient, respectively. (b) Superconducting transition temperature $T_C$ as a function of electron-phonon coupling constant $\lambda$ for bilayer $NbSe_2$. The solid line is the best fit to the strong-coupling formula (Eqn. 1) with $\omega_{log} = 50$ K and $\mu^* = 0.10$.

# Supplementary Information

## Gate tuning of electronic phase transitions in two-dimensional NbSe$_2$

Xiaoxiang Xi, Helmuth Berger, László Forró, Jie Shan*, and Kin Fai Mak*
Correspondence to: jus59@psu.edu, kzm11@psu.edu

## 1. Control experiments
### 1.1 Reversibility of the gating effects

We have investigated six bilayer NbSe$_2$ devices capped with monolayer MoS$_2$. All of them were reversible under ionic liquid gating within the gate range of -2 V to 3 V. (Some devices withstood higher gate voltages.) Supplementary figure 1 demonstrates the effect in one of the devices under the gating sequence of 0, -2, 0, 2, 0, 3, 0, and -3 V. We found both the sheet resistance and sheet Hall coefficient reproducible at zero gate voltage for the first three runs. The last run at 0 V (after applying 3 V) shows a hysteresis. The hysteresis is better seen in supplementary figure 2, which shows the sheet resistance at 220 K when the gate voltage was swept at 10 mV/s between -2.5 V and 2.5 V. The hysteresis, however, can be reversed if the device is kept at a negative gate voltage for a few minutes.

We note that at -3 V (pink lines in supplementary figure 1) the normal-state resistance starts to rise above the value at -2 V in the normal state although it is expected to be lower due to the enhanced electron-phonon interactions as discussed in the main text. We took this as an indication of the onset of sample degradations although its influence on the superconducting transition temperature $T_C$ may not be significant. Further applications of negative voltages beyond -3V cause irreversible sample degradations.

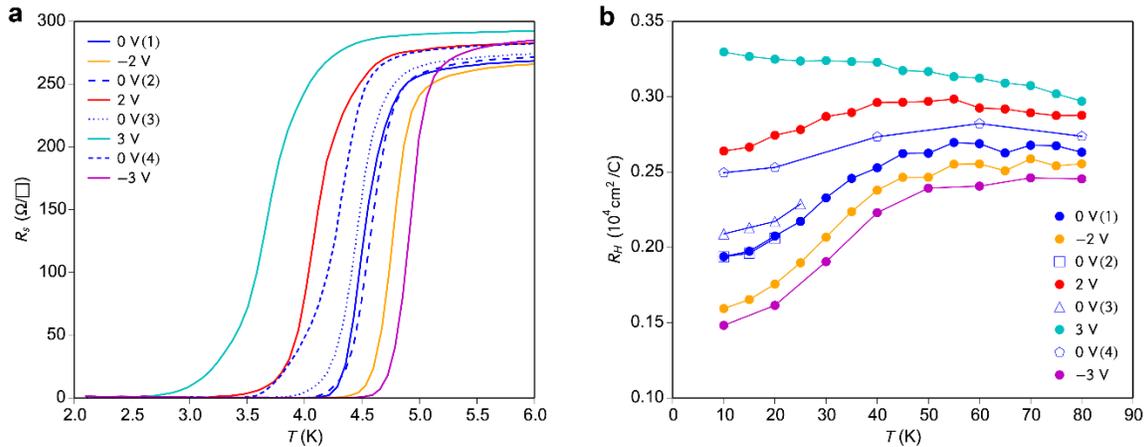

**Supplementary figure 1.** Temperature dependence of the sheet resistance (**a**) and the sheet Hall coefficient (**b**) of a bilayer NbSe$_2$ device capped with monolayer MoS$_2$. Different gate voltages were applied in the sequence of 0, -2, 0, 2, 0, 3, 0, and -3 V.

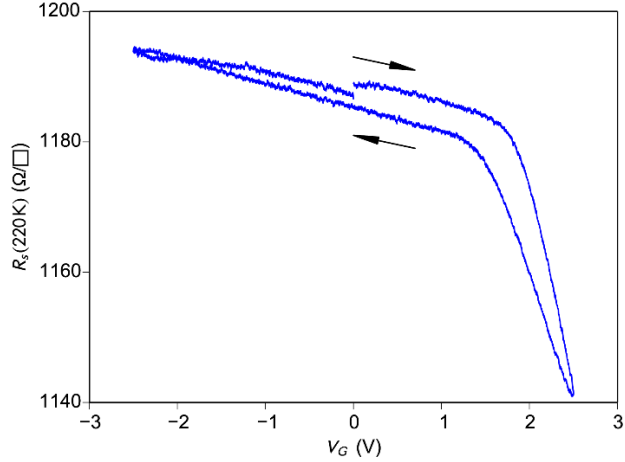

**Supplementary figure 2.** Gate dependence of the sheet resistance of bilayer $NbSe_2$ capped with monolayer $MoS_2$ measured at 220 K. The gate voltage was swept at 10 mV/s between -2.5 and 2.5 V.

### 1.2 Gating effects on the MoS$_2$ capping layer

To verify the gating effects on the $MoS_2$ capping layer in the electric-double-layer $NbSe_2$ device, we performed similar measurements on identical structures without the $NbSe_2$ bilayer. The device fabrication method is identical to what has been described in the main text. In this method, monolayer $MoS_2$ forms poor electrical contact with gold electrodes (often not conducting). Nevertheless, two-point resistance was measured for the conducting contacts at gate voltages from -2 V to 3 V (supplementary figure 3). The two-point resistance for both devices was well above $10^4$ ohms between 120 K and 220 K, nearly two orders of magnitude larger than that of the bilayer $NbSe_2$/monolayer $MoS_2$ devices. Measurements at lower temperatures were not possible since the contacts for both devices broke. Similarly, measurements were also not possible below ~ 200 K at a gate voltage of -2 V due to the contact problem.

In the second experiment, we fabricated an identical electric-double-layer device with bilayer $TaSe_2$ (instead of $NbSe_2$) capped by monolayer $MoS_2$ (supplementary figure 4a). $TaSe_2$ is an isoelectronic metal of $NbSe_2$ with $T_C$ ~ 0.2 K in the bulk [1]. The sheet resistance of the $TaSe_2$ device capped by $MoS_2$ shows no sign of superconductivity down to 2.1 K under ionic liquid gating up to $\pm$ 2 V (supplementary figure 4b). The kink in the resistance around 100 K is due to the CDW transition in $TaSe_2$ [2]. Combining these two experiments, we conclude that the observed superconductivity in bilayer $NbSe_2$ capped by monolayer $MoS_2$ around 5 K cannot be the gate-induced superconductivity in monolayer $MoS_2$; and monolayer $MoS_2$ serves barely as a protection dielectric layer in the structure. Further evidence is shown in the experiment discussed in the next section, where different capping materials produce similar gating effects in $NbSe_2$ devices.

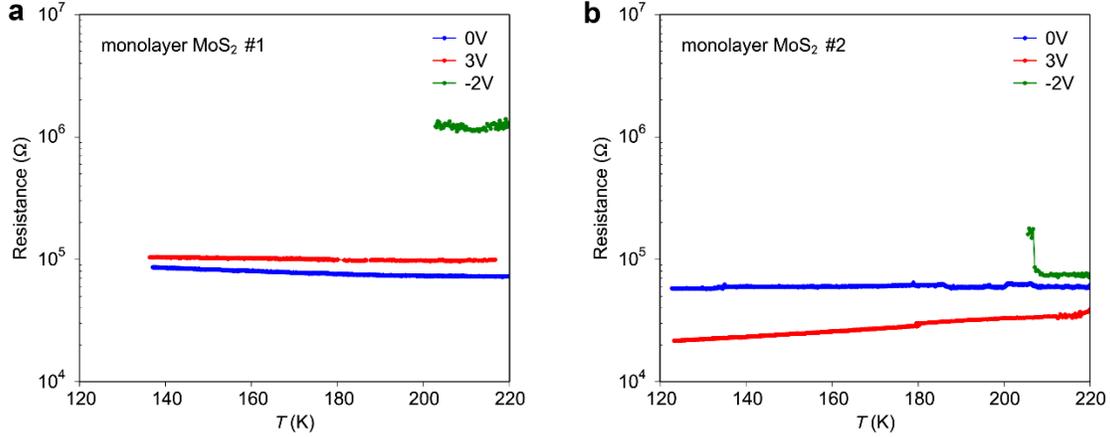

**Supplementary figure 3.** Temperature dependence of two-point resistance of two monolayer MoS$_2$ devices (**a** & **b**), with gate voltages applied in the sequence of 0, 3, and -2 V.

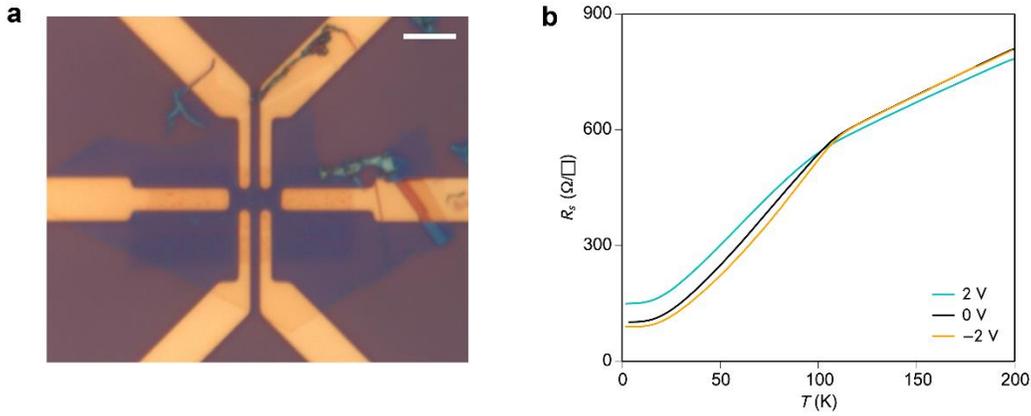

**Supplementary figure 4.** (**a**) Optical microscope image of a bilayer TaSe$_2$ device capped with monolayer MoS$_2$. Scale bar is 5 μm. (**b**) Temperature dependence of the sheet resistance of the bilayer TaSe$_2$ device at several selected gate voltages down to 2.1 K.

### 1.3 hBN-capped bilayer NbSe$_2$

Here we demonstrate that bilayer NbSe$_2$ capped by ultrathin hexagonal boron nitride (hBN) exhibits similar gate dependence of $T_C$ compared to that of bilayer NbSe$_2$ capped by monolayer MoS$_2$. The hBN capped bilayer NbSe$_2$ device was prepared and measured using the same methods as described in the main text. The hBN capping layer is ~ 1 nm thick, measured by atomic force microscopy. The results are shown in Supplementary figure 5. A $T_C$ modulation of 0.55 K was observed in the gate voltage range of $\pm$ 2 V, compared to 0.67 K for the same gate voltage range for a monolayer MoS$_2$ capped device. The reduced modulation in $T_C$ is likely due to the larger thickness of the hBN capping layer and the associated smaller gating efficiency. Significant gating effect was also observed in the temperature dependence of the sheet Hall coefficient, showing qualitatively the same behaviors as in devices capped by monolayer MoS$_2$. Indeed, as long as the Fermi level remains within the band gap of the capping material, gate-induced carrier density is expected to be at the NbSe$_2$ layers.

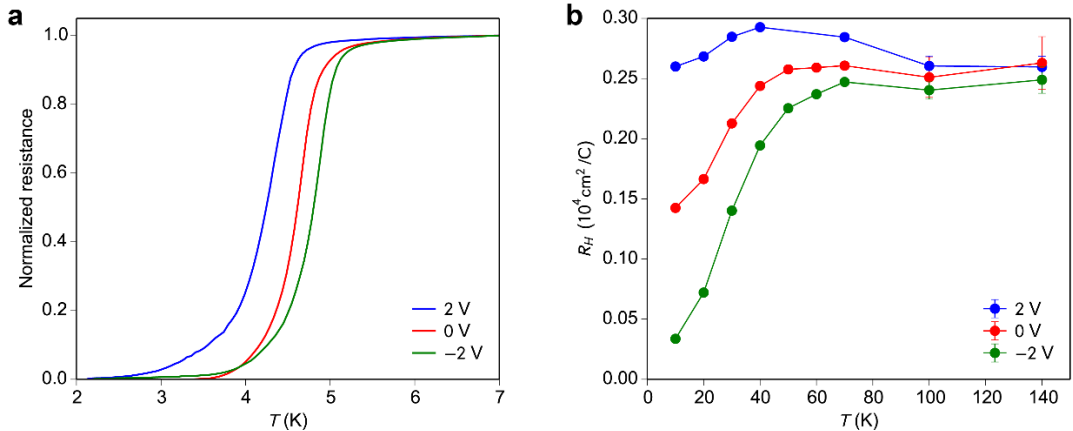

**Supplementary figure 5.** Electrical transport data for a bilayer NbSe$_2$ device capped with ultrathin hBN at different gate voltages. (**a**) Temperature dependence of the longitudinal sheet resistance normalized to its value at 7 K. (**b**) Temperature dependence of the sheet Hall coefficient.

### 1.4 Solid-state dual-gate bilayer NbSe$_2$ device

To compare the gating efficiencies, we have studied a dual-gate device with conventional solid-state dielectric gates. A schematic cross section of the device is shown in supplementary figure 6a and the device image in 6b. It consists of a vertical stack of bilayer NbSe$_2$/monolayer MoS$_2$/few-layer hBN/few-layer graphene on Si substrate with a 280-nm thermal oxide layer. The SiO$_2$/Si and the hBN/graphene serve, respectively, as the back and top gate. Bilayer NbSe$_2$ on Si substrate with pre-patterned electrodes was first prepared using the method described in the main text. The rest atomically thin layers (mechanically exfoliated from bulk crystals) were picked up using PPC (polypropylene carbonate) and transferred onto NbSe$_2$ one-by-one. The residual PPC was removed by anisole after each transfer. We were able to vary the back gate voltage from -100 to 100 V and the top gate voltage from -2 to 0.9 V without inducing significant leak currents. Combining the two gates we were able to tune $T_C$ by 0.16 K (supplementary figure 6c). This is < 4% variation in $T_C$ in contrast to the 50% variation achieved using ionic liquid gating.

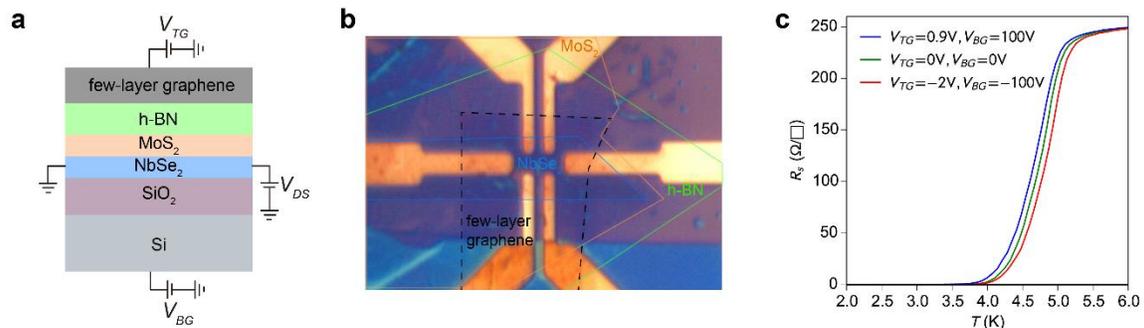

**Supplementary figure 6.** (**a**) Schematic cross section of a dual-gate device, showing the monolayer MoS$_2$/bilayer NbSe$_2$ stack sandwiched between a SiO$_2$/Si back gate and a hBN/ graphene top gate. (**b**)

Optical image of the device as shown schematically in (**a**). (**c**) Temperature dependence of the sheet resistance of bilayer NbSe₂ at selected gate voltages.

## 2. Hall resistance in bilayer NbSe₂

Supplementary figure 7 shows the magnetic field ($H$) dependence of the Hall resistance $R_{xy}$ of the device shown in figure 2 of the main text at different temperatures for a few gate voltages. We have obtained $R_{xy}(H) = \frac{R_t(H) - R_t(-H)}{2}$ by anti-symmetrizing the measured transverse resistance $R_t$ under magnetic field $H$ of two opposite out-of-plane directions due to the longitudinal-transverse coupling and the presence of magnetoresistance, which are symmetric in $H$ at low temperatures. The sheet Hall coefficient $R_H$ is obtained from the slope of the $R_{xy}(H)$ dependence divided by the bilayer thickness $t = 1.25$ nm [3].

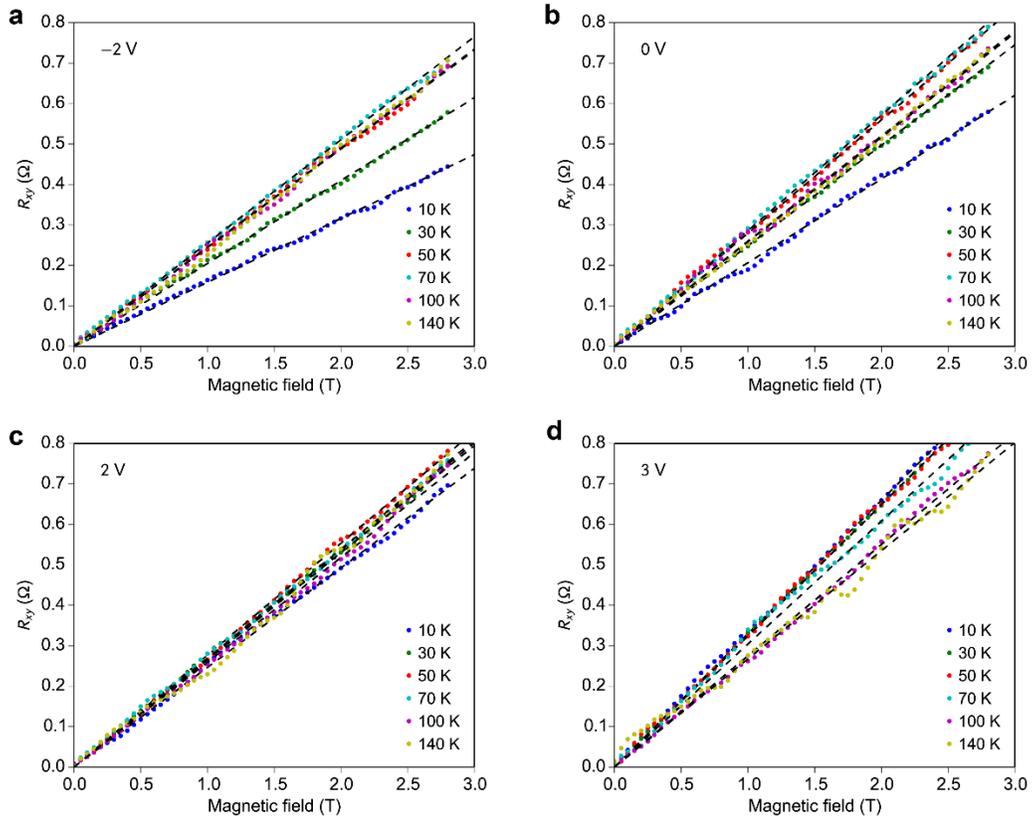

**Supplementary figure 7.** Magnetic field dependence of the Hall resistance of bilayer NbSe₂ at selected temperatures and gate voltages: -2 V (**a**), 0 V (**b**), 2 V (**c**) and 3 V (**d**). The dashed lines are linear fits.

## 3. Gate dependence of the mean-field and BKT transition temperature

In this section, we determine the Berezinskii-Kosterlitz-Thouless (BKT) transition temperature $T_\phi$ and the mean-field temperature $T_C$ for bilayer NbSe₂ at varying gate voltages. The BKT transition in two-dimensional (2D) superconductors corresponds to the disassociation of vortex-antivortex pairs [4]. The transition temperature $T_\phi$ can be determined from the power laws of the voltage-current dependence [5]. Supplementary figure 8a, b and c show such data at selected gate voltages for bilayer NbSe$_2$ capped with monolayer MoS$_2$. $T_\phi$ corresponds to the temperature at which $V \sim I^3$ (dashed lines). The gate dependence of $T_\phi$ is summarized in supplementary figure 8f.

The mean-field transition temperature $T_C$ of a dirty superconductor can be determined from the sheet resistance $R_S$ using the Aslamazov-Larkin formula [6] $R_S(T) = G_S^{-1} = \left(G_n + \frac{e^2}{16\hbar}\frac{T_C}{T-T_C}\right)^{-1}$ for $T \geq T_C$, where $G_s = 1/R_s$ is the sheet conductance and $G_n$ is the normal state sheet conductance. By taking the first derivative of the sheet conductance with respect to $T$, we rewrite the expression as

$$T - T_C = \sqrt{-\frac{e^2}{16\hbar}\left(\frac{dT}{dG_S}\right)T_C}. \tag{S1}$$

$T_C$ can thus be determined as the x-axis intercept of the $\sqrt{|dT/dG_s|}$ vs $T$ plot, as demonstrated in supplementary figure 8e. Its gate dependence is summarized in 8f. The figure also shows that $T_C$ at each gate voltage is very close to $T(0.5R_n)$. The ratio $T_\phi/T_C$ shows weak gate dependence. Its value is around 0.94.

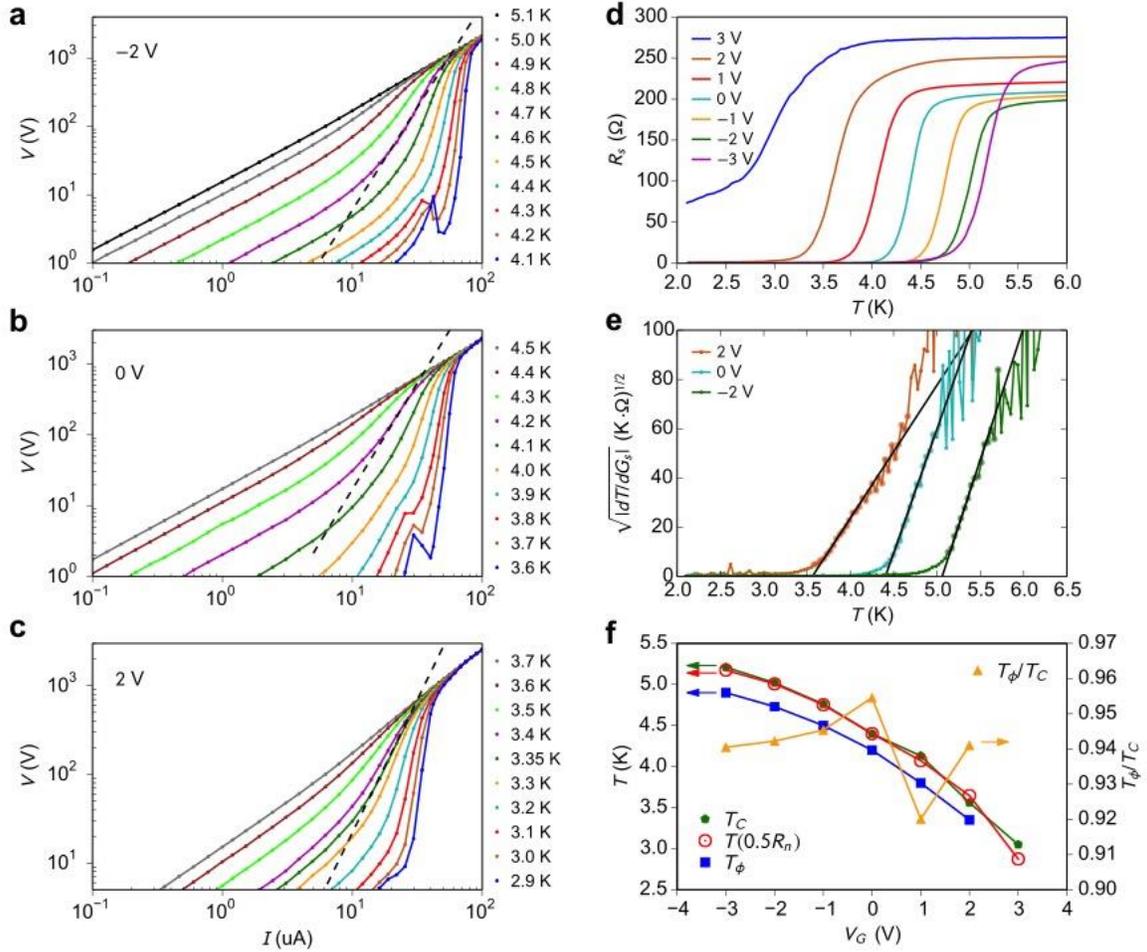

**Supplementary figure 8.** (**a-c**) Voltage vs current for bilayer NbSe$_2$ capped with monolayer MoS$_2$ at different temperatures, with the gate voltage at – 2 V (**a**), 0 V (**b**) and 2 V (**c**). The dashed lines indicate the $V \sim I^3$ dependence that is used to define $T_\phi$. (**d**) Temperature dependence of the sheet resistance across the superconducting transition at selected gate voltages for the same device. (**e**) $\sqrt{|dT/dG_s|}$ vs $T$ at 2 V, 0 V, and -2 V. The solid lines are linear fits to the data (symbols). (**f**) Gate voltage dependence of $T_C$, $T(0.5R_n)$ and $T_\phi$ (left axis) as well as the ratio $T_\phi/T_C$ (right axis).

## 4. Electron-phonon coupling constant extracted from sheet resistance

Here we use the electron-phonon scattering model to analyze the electron-phonon coupling constant of bilayer NbSe$_2$. In the high-temperature limit ($T/\Theta_D \gg 1$, where $\Theta_D$ is on the order of the Debye temperature), the bulk resistivity can be expressed as [7]

$$\rho \approx \rho_0 + \lambda \frac{2\pi k_B m}{\hbar e^2 n_0} T, \tag{S2}$$

where $\rho_0$ is the temperature independent resistivity due to electron-impurity scattering, $\lambda$ is the electron-phonon coupling constant, $m$ is the effective mass, and $n_0$ is the volume carrier density. For bilayer NbSe$_2$ of thickness $t$, the sheet resistance and sheet carrier density become $R_S = \rho/t$ and $n_{2D} = n_0 \cdot t$, respectively. Therefore

$$R_S(T) \approx R_0 + \lambda \frac{2\pi k_B m}{\hbar e^2 n_{2D}} T = R_0 + \lambda \frac{2\pi \hbar n_0 k_B}{\varepsilon_0 n_{2D}(\hbar \omega_{p0})^2} T. \tag{S3}$$

In the last step, we have introduced the plasma frequency $\omega_p = \sqrt{n_0 e^2/\varepsilon_0 m}$ ($\varepsilon_0$ the vacuum permittivity) of the bulk material ($\hbar \omega_p = 2.74$ eV for bulk NbSe$_2$ [8]). We fitted the temperature dependence of the sheet resistance of bilayer NbSe$_2$ between 65 – 180 K using eqn. S3 (supplementary figure 9a-c). For instance, at zero gate voltage, fit using eqn. S3 yields a slope of $dR_S/dT = 5.08$ Ohm/K. The electron-phonon coupling constant can then be calculated from the slope to be 1.2. The gate dependence of the electron-phonon coupling constant is shown in supplementary figure 9d.

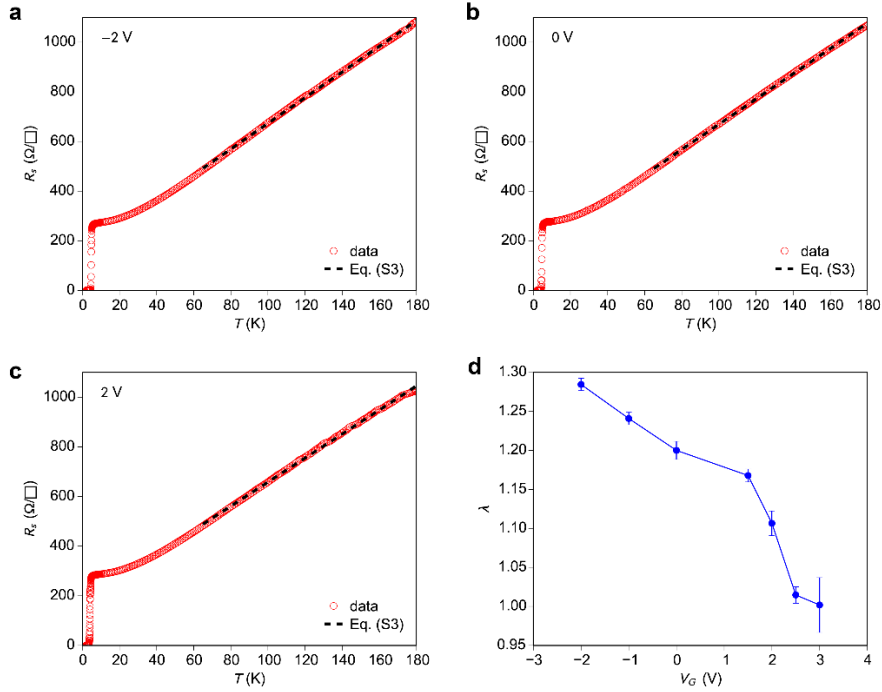

**Supplementary figure 9.** (**a-c**) Temperature dependence of the sheet resistance of bilayer NbSe$_2$ at gate voltage -2 V (**a**), 0 V (**b**), and 2V (**c**) (symbols). Dashed lines are fits to Equation (S3) between 65 – 180 K. (**d**) Extracted electron-phonon coupling constant as a function of gate voltage.

## Supplementary references

[1]    T. Kumakura, H. Tan, T. Handa, M. Morishita, and H. Fukuyama, Czech J Phys **46**, 2611 (1996).
[2]    M. Naito and S. Tanaka, Journal of the Physical Society of Japan **51**, 219 (1982).
[3]    J. A. Wilson, F. J. Di Salvo, and S. Mahajan, Adv Phys **50**, 1171 (2001).
[4]    M. R. Beasley, J. E. Mooij, and T. P. Orlando, Physical Review Letters **42**, 1165 (1979).
[5]    M. Tinkham, *Introduction to superconductivity* (McGraw-Hill, United States of America, 1996), 2nd edn.
[6]    L. G. Aslamasov and A. I. Larkin, Physics Letters A **26**, 238 (1968).
[7]    G. Grimvall, *The Electron-Phonon Interaction in Metals* (North-Holland Publishing Co., 1981).
[8]    S. V. Dordevic, D. N. Basov, R. C. Dynes, and E. Bucher, Physical Review B **64**, 161103 (2001).